\documentclass[conference]{IEEEtran}
\usepackage{amsmath}
\usepackage{amssymb}
\usepackage{datetime}
\usepackage{graphicx}

\ifCLASSINFOpdf
\else
\fi
\begin{document}
\setlength{\parskip}{0cm}
\title{Propagation of Uncertainty and Analysis of Signal-to-Noise in Nonlinear Compliance
Estimations of an Arterial System Model}

\author{\IEEEauthorblockN{Timothy S. Phan}
\IEEEauthorblockA{Department of Electrical and Computer Engineering\\
Rutgers University\\
Piscataway, NJ 08854, USA\\
tsphan@rutgers.edu}
\and
\IEEEauthorblockN{John K-J. Li}
\IEEEauthorblockA{Department of Biomedical Engineering\\
Rutgers University\\
Piscataway, NJ 08854, USA\\
johnkjli@rci.rutgers.edu}
}

\maketitle

\begin{abstract}
The arterial system dynamically loads the heart through changes in arterial compliance. The pressure-volume relation of arteries is known to be nonlinear, but arterial compliance is often modeled as a constant value, due to ease of estimation and interpretation. Incorporating nonlinear arterial compliance affords insight into the continuous variations of arterial compliance in a cardiac cycle and its effects on the heart, as the arterial system is coupled with the left ventricle. We recently proposed a method for estimating nonlinear compliance parameters that yielded good results under various vasoactive states. This study examines the performance of the proposed method by quantifying the uncertainty of the method in the presence of noise and propagating the uncertainty through the system model to analyze its effects on model predictions. Kernel density estimation used within a bootstrap Monte Carlo simulation showed the method to be stable for various vasoactive states.
\end{abstract}


\IEEEpeerreviewmaketitle

\section{Introduction}
Arterial pressure waveforms are manifestations of the coupled interactions between the heart and the arterial system. The left ventricle ejects its stroke volume into the aorta during systole when the aortic valve is open and decouples from the arterial system in diastole when the aortic valve is closed. This pulsatile nature of the pumping function of the heart is able to lead to steady perfusion of the body's organs due to properties of the arterial system. The complex load presented by the arterial system that permits the conversion from intermittent blood ejection to steady organ perfusion can be adequately represented by a modified 3-element Windkessel model \cite{westerhof2009arterial}, as judged by the close correspondence of the model's input impedance to the measured input impedance of the arterial system.

\begin{figure}[!t]
    \centering
    \includegraphics[width=3.5in]{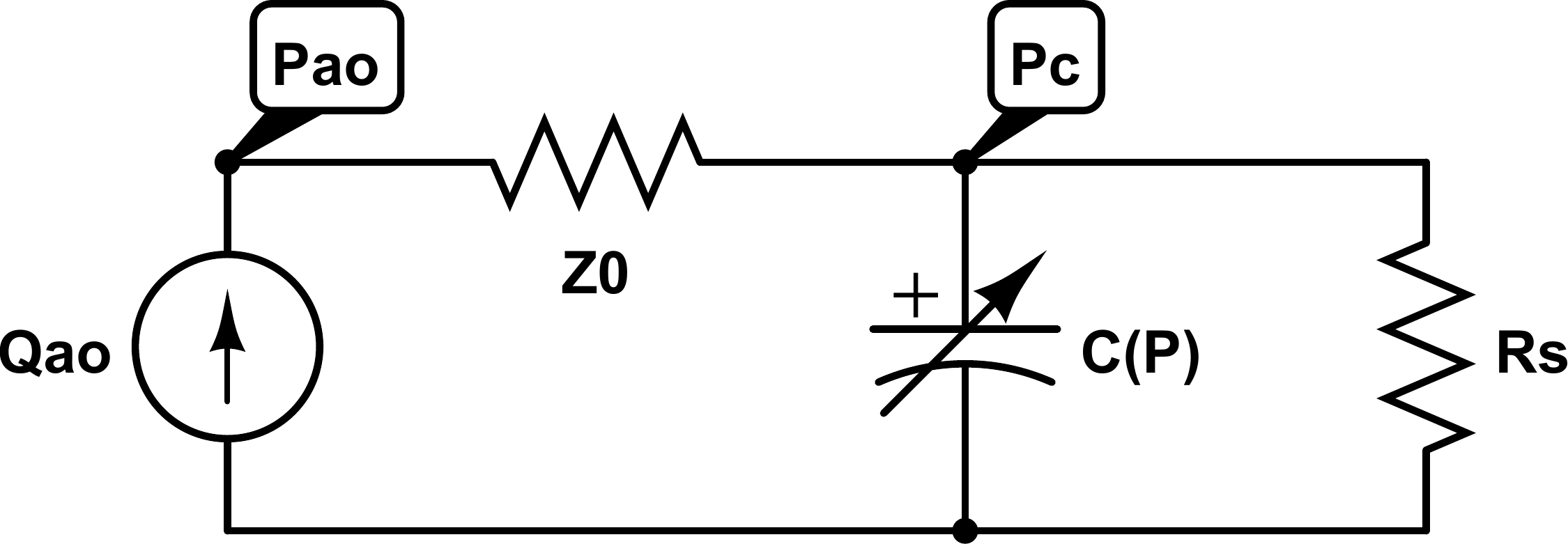}
    \caption{Modified windkessel with nonlinear pressure-dependent compliance.}
    \label{wind_fig}
\end{figure}

Impedance matching of the heart to the arterial system is important for minimizing wasteful reflections of pressure and flow waves generated by left ventricle. This coupling between the two systems permits gaining insight into the loading effect of the arterial system on the heart by analysis of arterial properties. Typically, arterial compliance is modeled as a constant value element, for instance, as an ordinary ideal capacitor in the Windkessel model shown in Fig. \ref{wind_fig}, although it is known that the pressure-volume relation of  arteries is nonlinear \cite{li1990nonlinear}; this may be attributed to difficulties in estimating parameters of nonlinear compliance from limited data. Incorporating nonlinear compliance into arterial system models permits a view of the continuous compliance variations throughout a cardiac cycle. The value of this added description is that due to the ``compliance matching'' of the arterial system to the left ventricle, this additional information provides insight into how the heart copes with various forms of arterial loading as the compliance of the left ventricle follows a complementing variation \cite{li1994arterial}. Furthermore, the differential effects of pharmacological interventions on passive and active modifications of the arterial wall can be studied by analyzing how nonlinear parameters change.

Nonlinear compliance parameters have generally been estimated using iterative methods that seek out suitable sets of parameters in parameter-space that minimize discrepancies between model-predicted and measured pressure waveforms \cite{matonick2001pressure}. Some have proposed methods that estimate the parameters by deriving pressure-volume curves from aortic flow and pressure measurements, but found limited success in vasodilation states and systems with low pressures \cite{cappello1995identification}. We recently proposed a new method that provides good estimates for normotensive, vasoconstriction, and vasodilation states representing normal pressure, high pressure, and low pressure systems, respectively \cite{phan2013reduced}. The present investigation examines the stability of the proposed method in the presence of noise and the effects of uncertainty in parameter estimates on model-predicted waveforms.

\section{Methods}

\subsection{Experimental}
Aortic flow ($Q_{ao}$) with an electromagnetic flow probe and
aortic pressure ($P_{ao}$) with a catheter-tip pressure
transducer were simultaneously recorded together with lead-II
ECG in anesthetized, ventilated mongrel dogs. Both pressure and
flow transducers were calibrated and tested to have adequate
frequency response. Vasoactive states were altered with
intravenous infusion of methoxamine to induce vasoconstriction,
and with nitroprusside to cause vasodilation. Data were recorded
and later digitized at 10 msec intervals for further analysis.
Experimental protocols were approved by the Rutgers University
Institutional Animal Care and Use Committee (IACUC).

\subsection{Arterial System Model}
A modified Windkessel model incorporating nonlinear pressure-dependent compliance, Fig. \ref{wind_fig}, was used for this study \cite{li1990nonlinear}. The pressure across and flow through the compliant element, represented by the variable capacitor, are
	\begin{equation}
	    \label{pc}
		P_c(t) = P_{ao}(t) - Q_{ao}(t)Z_0
	\end{equation}
	\begin{equation}
	    \label{qc}
		Q_c(t) =  Q_{ao}(t) - \frac{P_{ao}(t)-Q_{ao}(t)Z_0}{R_s} .
	\end{equation}

\begin{figure}
    \centering
    \includegraphics[width=3.5in]{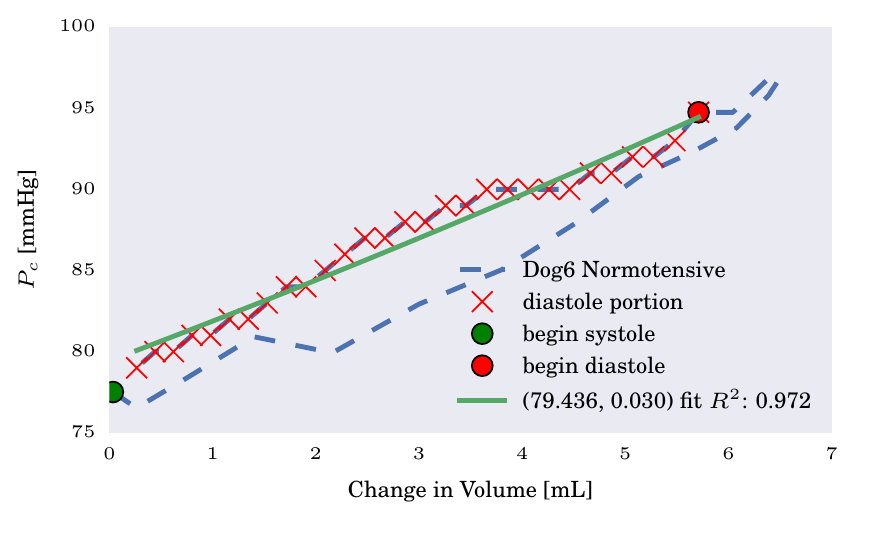}
    \caption{Pressure-volume curve with diastolic portion fitted to $P_c(V_c) = a_p e^{b_pV_c}$. Estimated parameters are shown in parenthesis.}
	\label{pv_fit}
\end{figure}
	
  Peripheral resistance, representing the steady load to the heart, is calculated as the mean aortic pressure divided by mean aortic flow,
    \begin{equation}
        R_s = \frac{\bar{P}_{ao}}{\bar{Q}_{ao}}.
    \end{equation}

  The characteristic impedance of the aorta, representing the aortic pressure-flow relationship in the absence of reflections, can be calculated as
    \begin{equation}
        Z_0 = \frac{P_{ao}(t)-P_{diastole}}{Q_{ao}(t)}
    \end{equation}
    during the first 60-80ms of systole, when reflections are assumed to be small \cite{li1989increased}.

Compliance of arteries is defined as arterial volume variations with respect to varying arterial pressure.  Nonlinear pressure-dependent compliance of this model is exponentially related to the pressure across the compliant element,
	\begin{equation}
		\label{c_pc}
		C(P_c) \triangleq \frac{dV_c}{dP_c} = ae^{-bP_c}.
	\end{equation}

A method recently proposed for estimating the parameters of nonlinear compliance is summarized here \cite{phan2013reduced}.

Integrating $C(P_c)$ with respect to pressure yields the functional forms of pressure-volume,
	\begin{equation}
	    \label{pv_orig}
		V_c(P_c) = \frac{-ae^{-bP_c}}{b} + K \Rightarrow P_c(V_c) = \frac{1}{b}\left[ ln\left(\frac{a}{b(K-V_c)}\right)\right] .
	\end{equation}

  Mass conservation allows estimation of arterial volume change relative to end-diastolic volume, $V_{ed}$, as a function of flow through the branch containing the compliant element,
	\begin{equation}
		\label{v_qc}
		\Delta V_c(t) = V_c(t)-V_c(t_{ed}) = V_c(t) - V_{ed} = \int_{t_{ed}}^{t} Q_c(\tau) d\tau .
	\end{equation}

    Each point of $\Delta V_c$ can be related in time to a corresponding $P_c$, yielding a pressure-volume curve. As the aortic valve is closed during diastole, decoupling the heart from the arterial system, fitting \eqref{pv_orig} to the diastolic portion of the pressure-volume curve removes cardiac influences \cite{stergiopulos1995evaluation}. The nonlinear regression required to estimate the three parameters, \{$a,b,K$\}, yields limited convergence in vasodilation states and systems with low pressure \cite{cappello1995identification}.
\begin{figure}[!t]
    \centering
    \includegraphics[width=3.5in]{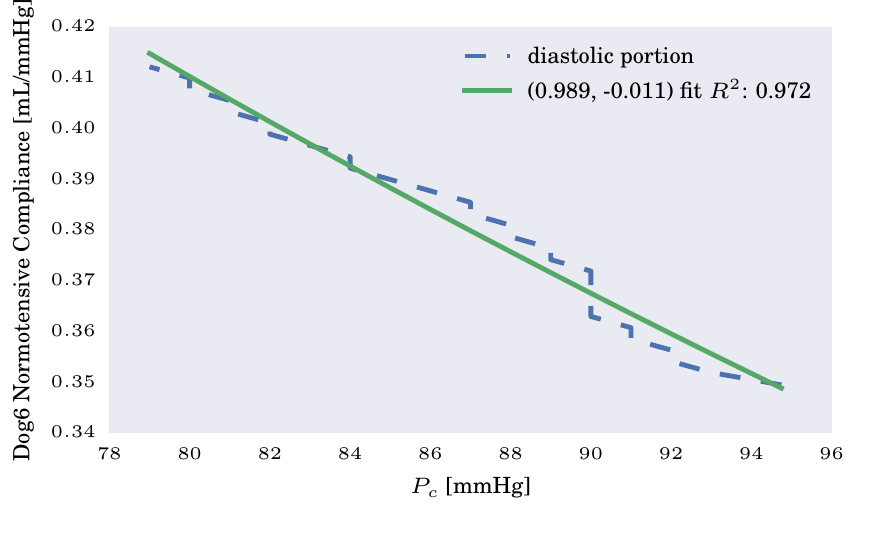}
    \caption{Compliance-pressure curve during diastole for a normotensive dog. Fitted parameters are shown in parenthesis.}
	\label{cp_fit}
\end{figure}

Assuming that arteries are pre-stressed to diastolic pressure $P_d$, the effective origin of their exponential pressure-volume curve is located at $P_d$ so that $P_c$ can be locally approximated to a 2-parameter form as
	\begin{equation}
		\label{pv_app}
		P_c(V_c) \approx a_p e^{b_pV_c} = P_d e^{b_pV_c} .
	\end{equation}
This approximation allows reduction of the regression problem to primarily estimating one parameter, $b_p$, which was found to be tractable in normotensive, vasoconstriction, and vasodilation states representing normal pressure, high pressure, and low pressure systems, respectively.
	
	The modulus of volume elasticity, E, defined as the inverse of volume compliance, is thus approximated as
	\begin{equation}
		\label{elasticity}
		E \triangleq \frac{dP_c}{dV_c} = \frac{1}{C} \approx b_pP_de^{b_pV_c} .
	\end{equation}

The estimated $b_p$ can be used to analytically calculate $C = E^{-1}$ using values of $\Delta V_c$ calculated from \eqref{v_qc}. A point-by-point plot of $C$ against its corresponding $P_c$ in time yields a curve that can be fitted to \eqref{c_pc} for estimates of $a$ and $b$.

	With estimations of the nonlinear compliance parameters, the predicted aortic pressure from the modified Windkessel model of Fig. \ref{wind_fig} with aortic flow as input can be obtained using methods from electrical circuit theory to solve for pressure at the aortic node. Retaining the values of compliance at each integration step allows production of compliance variation with time over the cardiac cycle.

\subsection{Uncertainty Model}
Pressure-volume curves generated with \eqref{v_qc} will inevitably contain noise from experimental measurements of aortic flow and pressure (\textit{measurement noise}). As models are inherently simplifications of reality, model-predicted values may not exactly match experimentally measured values (\textit{model noise}), which will also include errors from numerical integration. Consequently, the \textit{i-th} residual, $\hat{\epsilon_i}$, from fitting \eqref{pv_app} to the pressure-volume curve will contain both measurement and model noise, here modeled as
	\begin{equation}
		\hat{\epsilon_i} = P_{c_i} - \hat{P}_{c_i} = N_{meas_i} + N_{mod_i} .
	\end{equation}

\begin{figure}[!t]
    \centering
    \includegraphics[width=3.5in]{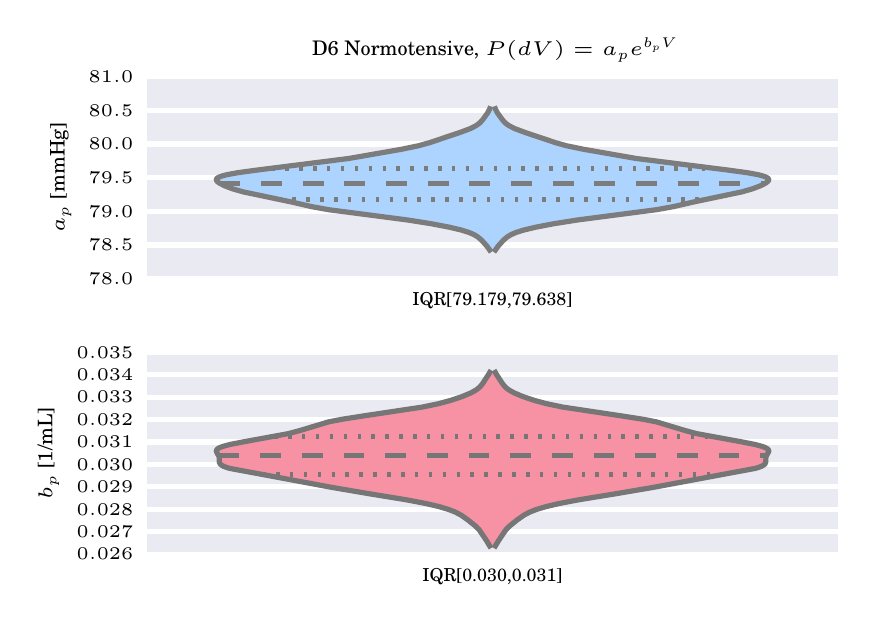}
    \caption{Modified boxplot showing distributions of normotensive pressure-volume parameters obtained from Monte Carlo simulations. Interquartile range is bounded by the dotted (...) lines. The median is represented by the dashed (---) lines.}
    \label{violin_npv}
\end{figure}

Kernel density estimation (KDE), a non-parametric density estimation technique, was used to obtain a smooth estimate of the probability density function (pdf) of $\hat{\epsilon}$. This is to avoid imposing a particular (e.g. normal) distribution on noise terms, allowing for a more flexible analysis of uncertainty by bootstrapping the residuals for Monte Carlo assessment of uncertainty. Gaussian kernels with bandwidth selected using Scott's Rule was used in the KDE process \cite{scott2009multivariate}.
\begin{table}[!h]
\renewcommand{\arraystretch}{1.3}
\caption{Signal-to-noise ratios for estimated pressure-volume and compliance-pressure parameters.}
\label{snr_table}
\centering
\begin{tabular}{c|c|c|c}
	\hline
	& \textbf{Normotensive}  &  \textbf{Hypertensive} & \textbf{Vasodilated}\\
	\hline
	\hline

	$a_p$(mmHg)	&   233.8		&  147.1	&  263.1\\
	\hline

	$b_p$(1/mL)	&   24.49  	&  14.10	&  23.36\\
	\hline

 	$a$(mL/mmHg)		& 	29.83 	&  14.41	&  25.06\\
	\hline

	$b$(1/mmHg)		&	28.40	&  13.26	&  24.42\\
	\hline
\end{tabular}
\end{table}
A Monte Carlo procedure was implemented to generate $N=1000$ virtual pressure-volume curves by adding noise sampled from the pdf of $\hat{\epsilon}$ to an initial fit of \eqref{pv_app} for the pressure-volume curve,
	\begin{equation}
		\hat{P_c}(\hat{V}_c) = P_{d_0} e^{b_{p_0}\hat{V}_c} + \hat{\epsilon}_p(\hat{V}_c),
	\end{equation}
where \{$P_{d_0},b_{p_0}$\} is the set of parameters from the initial fit, $\hat{\epsilon}_p(\hat{V}_c)$ is a point sampled from the pdf representing the noise, and $\hat{P_c}(\hat{V}_c)$ is a point on the generated virtual pressure-volume curve.  Each of the virtual curves was then fitted to \eqref{pv_app} and the resulting N estimates of \{$\hat{P}_d,\hat{b}_p$\} were stored in an array for analysis. 

The initial pressure-volume parameter set \{$P_{d_0},b_{p_0}$\} was then used to generate a $C$ vs. $P_c$ plot as previously described, and the resulting curve was fitted to the nonlinear pressure-dependent compliance form of \eqref{c_pc} to obtain an initial nonlinear compliance parameter set \{$a_{0},b_{0}$\}. Residuals from the fit of the compliance-pressure curve are similarly modeled to be the result of measurement and model noise,
	\begin{equation}
		\hat{C}(\hat{P}_c) = a_0 e^{-b_0 \hat{P}_c} + \hat{\epsilon}_c(\hat{P}_c).
	\end{equation}

The same KDE and Monte Carlo procedures were used to estimate the pdf of the residuals from the initial fit, sample noise from the estimated noise pdf, and generate virtual compliance-pressure curves to obtain an array of N sets of nonlinear compliance parameters \{$\hat{a},\hat{b}$\}.
\begin{figure}[!t]
    \centering
    \includegraphics[width=3.5in]{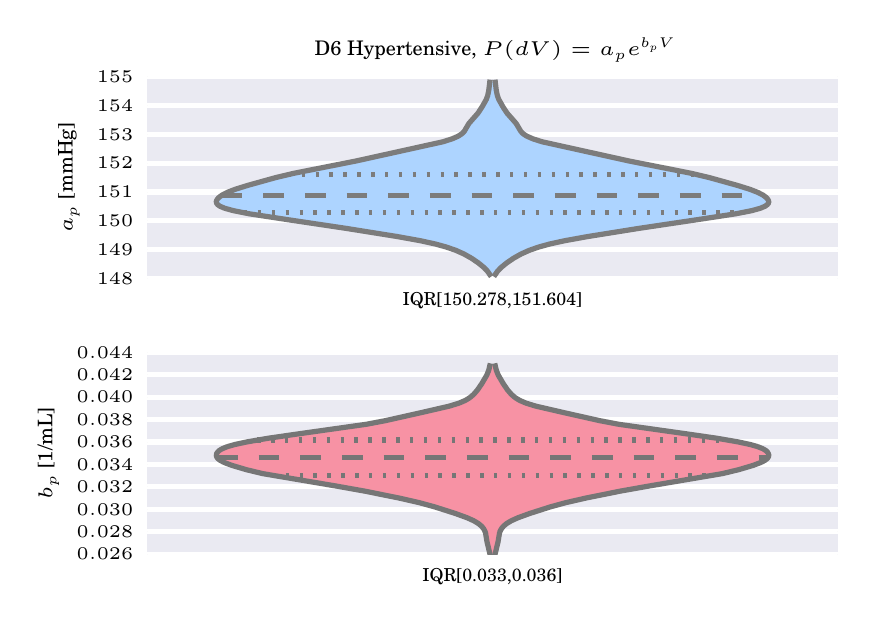}
    \caption{Modified boxplot for hypertensive pressure-volume parameters. Interquartile range is bounded by the dotted (...) lines. The median is represented by the dashed (---) lines.}
    \label{violin_hpv}
\end{figure}

Signal-to-noise ratio is defined as
	\begin{equation}
		SNR = \frac{\mu}{\sigma},
	\end{equation}
where $\mu$ and $\sigma$ are the mean and standard deviation, respectively, of a particular parameter (e.g. $b_p$). This definition was used as a measure of dispersion and stability on estimates of the parameters. A KDE of each array of parameters was performed to analyze how the estimated parameters were distributed in the presence of uncertainty.

Uncertainty in the nonlinear compliance parameter estimation was propagated through the arterial system model by using the N sets of \{$a,b$\} to generate N model-predicted aortic pressure and compliance waveforms.

\section{Results}

\begin{figure}[!t]
    \centering
    \includegraphics[width=3.5in]{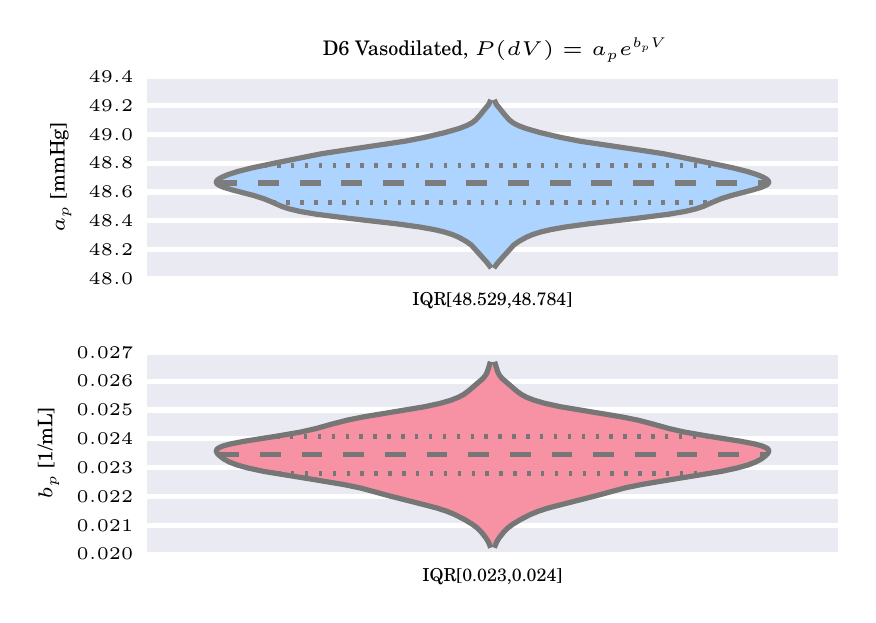}
    \caption{Modified boxplot for vasodilated pressure-volume parameters. Interquartile range is bounded by the dotted (...) lines. The median is represented by the dashed (---) lines.}
    \label{violin_vpv}
\end{figure}

	Representative pressure-volume and compliance-pressure curves used for estimation of nonlinear arterial compliance estimations are shown in Figs. \ref{pv_fit} and \ref{cp_fit}. The coefficient of determination, $R^2$, for both fits is 0.972 and the plots show good agreement between the data and proposed model. It can be seen that $a_p$ for this condition is 79.436 mmHg, which corresponds to the diastolic pressure of this particular cardiac cycle. The estimation for $b_p$, representing the volume-dependence of pressure, is 0.030 mL$^{-1}$. Use of \eqref{elasticity} generates the compliance-pressure curve in Fig. \ref{cp_fit}. $a$ and $b$ are estimated to be 0.989 mL/mmHg and 0.011 mmHg$^{-1}$, respectively.

	KDE of the residuals from pressure-volume and compliance-pressure fits were used in the Monte Carlo procedure to produce a distribution of parameters. Figs. \ref{violin_npv}-\ref{violin_vpv} show modified boxplots of the estimated pressure-volume parameters; the sides of the boxplots are formed by KDE of the pressure-volume parameters, which show the distributions resembling a normal distribution. The dotted lines within the boxplots encompass the interquartile range of the distributions. Normotensive and vasodilated states had interquartile range of 0.001 for $b_p$, whereas hypertensive had an interquartile range of 0.003.

\begin{figure}[!t]
    \centering
    \includegraphics[width=3.5in]{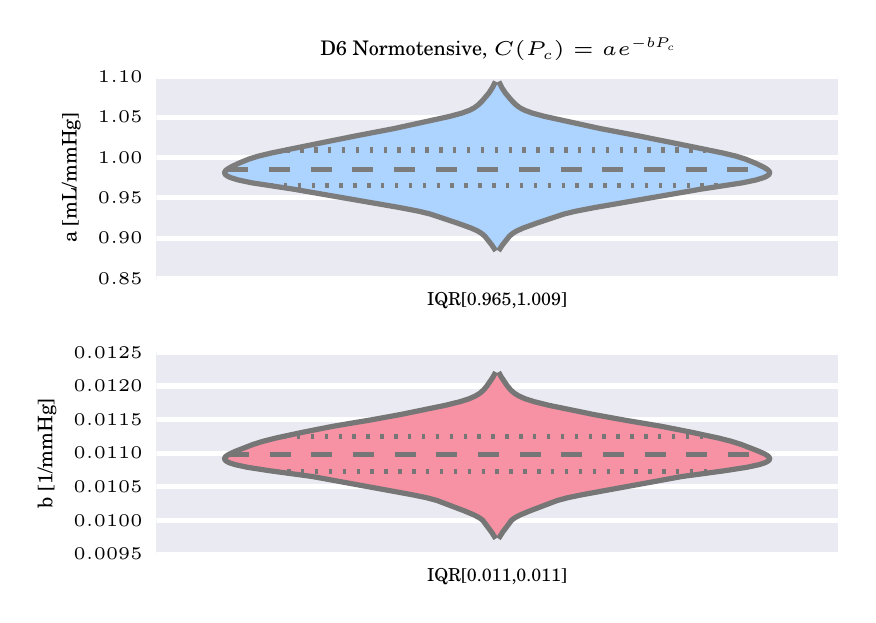}
    \caption{Modified boxplot showing distributions of normotensive compliance-pressure parameters obtained from Monte Carlo simulations. Interquartile range is bounded by the dotted (...) lines. The median is represented by the dashed (- - -) lines.}
    \label{violin_ncp}
\end{figure}

\begin{figure}[!h]
    \centering
    \includegraphics[width=3.5in]{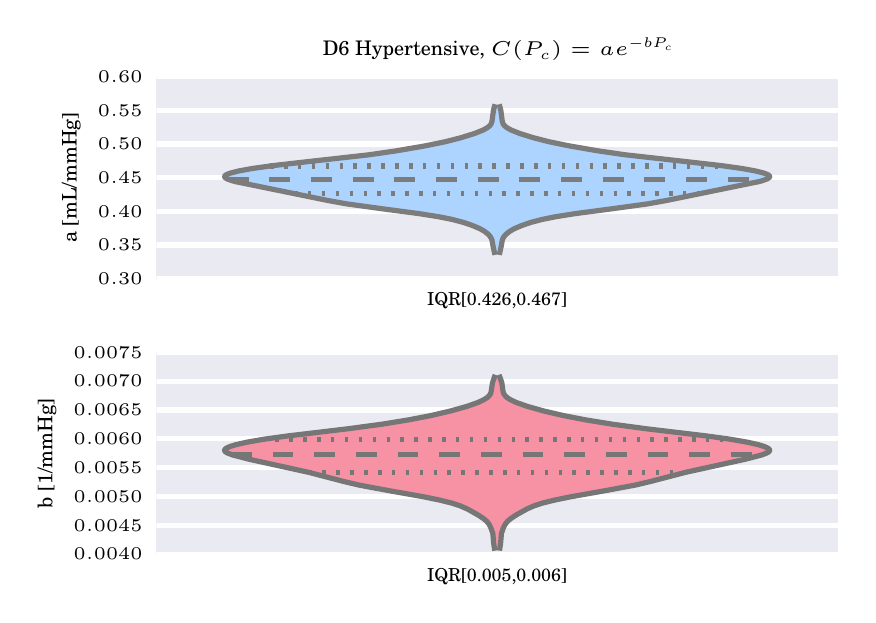}
    \caption{Modified boxplot for hypertensive pressure-volume parameters. Interquartile range is bounded by the dotted (...) lines. The median is represented by the dashed (- - -) lines.}
    \label{violin_hcp}
\end{figure}

	The distribution of estimated compliance parameters are shown in the modified boxplots of Figs. \ref{violin_ncp}-\ref{violin_vcp}. All distributions resemble a normal distribution. Interquartile ranges show narrow ranges for estimated compliance parameters in all the vasoactive states. Signal-to-noise ratios are shown in Table \ref{snr_table}. Vasodilated and normotensive states displayed higher signal-to-noise ratios than the hypertensive state, for all estimated parameters.

\begin{figure}[!t]
    \centering
    \includegraphics[width=3.5in]{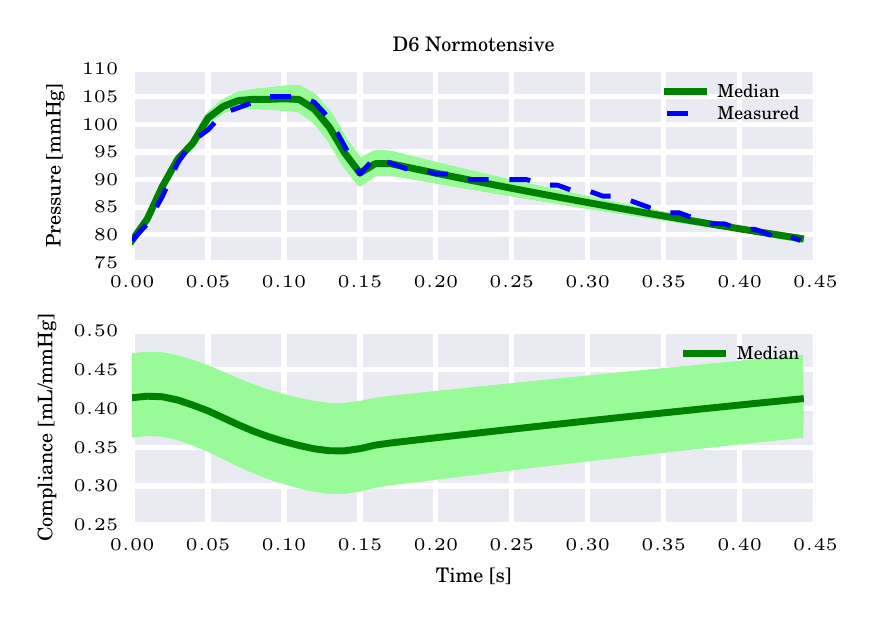}
    \caption{Comparison of normotensive measured and model-predicted aortic pressure and compliance waveforms after propagating uncertainties in nonlinear arterial compliance parameters. Thick green lines represent the median predicted waveforms while the shaded green regions indicate the Monte Carlo 95\% confidence intervals.}
    \label{normo}
\end{figure}

\begin{figure}[!h]
    \centering
    \includegraphics[width=3.5in]{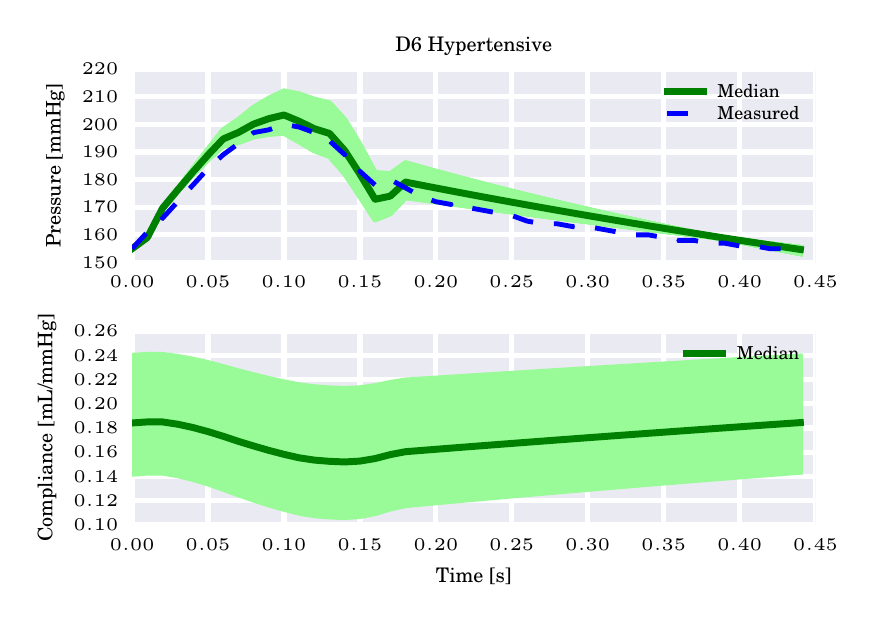}
    \caption{Propagated uncertainties for hypertensive state. Thick green lines represent the median predicted waveforms while the shaded green regions indicate the Monte Carlo 95\% confidence intervals.}
    \label{hyper}
\end{figure}

\begin{figure}[!h]
    \centering
    \includegraphics[width=3.5in]{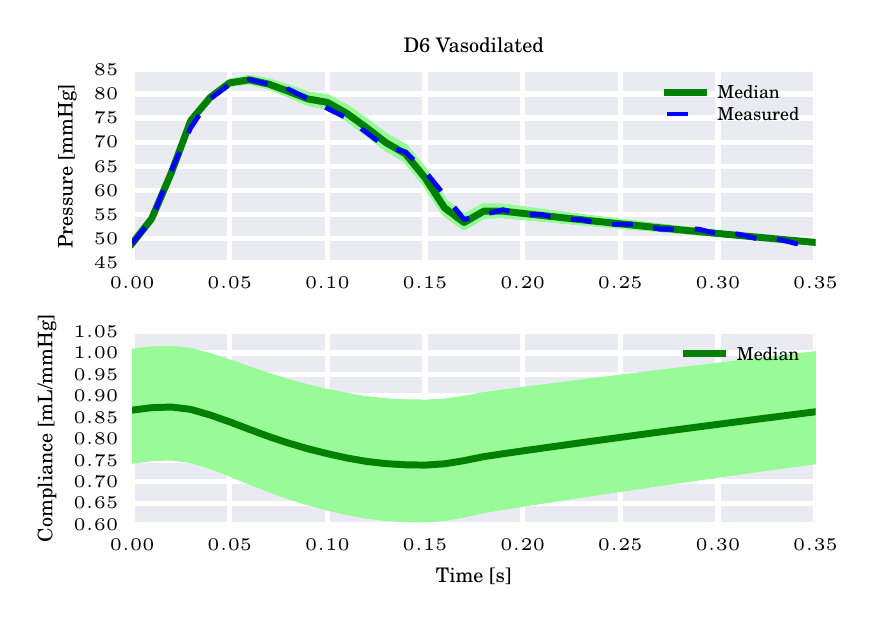}
    \caption{Propagated uncertainties for vasodilated state. Thick green lines represent the median predicted waveforms while the shaded green regions indicate the Monte Carlo 95\% confidence intervals.}
    \label{vaso}
\end{figure}

	When all sets of nonlinear compliance parameters from the Monte Carlo simulations were propagated through the arterial system model, the most prominent effects were on the systolic peaks, as seen in Figs. \ref{normo}-\ref{vaso}. The shaded green areas of the pressure and compliance time-series represent the Monte Carlo 95\% confidence intervals based upon uncertainties of compliance parameters and the thick green line represents the median. Comparing the confidence intervals of the compliance plots across vasoactive states show no overlap. The vasodilated state shows greatest magnitude of variation throughout a cardiac cycle whereas the hypertensive state shows least amount of variation (i.e. the hypertensive arteries appear stiff). In all vasoactive states, the 95\% confidence interval for model-predicted aortic pressure adequately captures the actual measured pressure; the median aortic pressure curve matches very closely to the curve of the measured pressure.

\section{Discussion}

\begin{figure}
    \centering
    \includegraphics[width=3.5in]{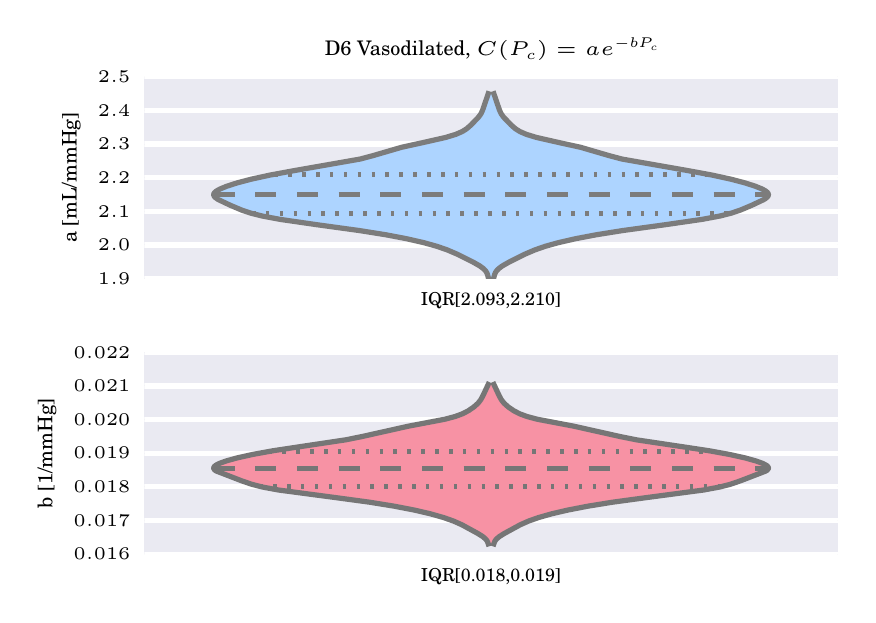}
    \caption{Modified boxplot for vasodilated pressure-volume parameters. Interquartile range is bounded by the dotted (...) lines. The median is represented by the dashed (- - -) lines.}
    \label{violin_vcp}
\end{figure}

The narrow ranges for parameter estimates in the presence of noise, as shown in the various modified boxplots of Figs. \ref{violin_npv}-\ref{violin_vcp}, provide support that in the presence of measurement and model noise, as well as under various vasoactive states, the proposed method converges to a stable set of parameters. Signal-to-noise ratios were used to quantitatively compare dispersions of parameter estimates across vasoactive states. Since the method presented in \cite{cappello1995identification} failed to obtain good estimates in normotensive and vasodilated states, it was interesting to observe from the signal-to-noise ratios that our proposed method showed higher ratios in those two states as compared to the hypertensive state. This suggests that the robustness of our method across vasoactive states may be attributed to the reduced-order nature of our method.

It can be seen that the signal-to-noise ratio for the pressure-volume parameter $b_p$ is directly related to the ratios for both compliance-pressure parameters. That is, under a particular vasoactive state, the magnitude of the signal-to-noise of the pressure-volume parameter $b_p$ appears to dictate the level of uncertainty in nonlinear compliance parameters. Since the parameter $a_p$ is easily known a priori with our method as it is equal to diastolic pressure of the cardiac cycle under study, and the burden of the method primarily rests upon stable estimates of $b_p$, it is of interest to further investigate the performance of our method in the presence of reflected waves that more heavily distort the diastolic portion of the pressure waveform.  The advantage of using Monte Carlo simulations that involve bootstrapping the residuals of the proposed estimation method is that the process allows use of the proposed estimator itself to quantify its own uncertainty, instead of making assumptions of normally distributed errors.

Propagating the uncertainty in compliance parameters causes widening of the confidence interval around the systolic peak of the model-predicted aortic pressure waveforms. This is consistent with the fact that arterial compliance, as the dynamic load to the heart, alters pulse pressure, whereas peripheral resistance, as the steady load to the heart, causes shifts in the mean arterial pressure. It is encouraging to see that in the presence of noise and the uncertainties in parameter estimates of our method, the 95\% confidence intervals of the predicted pressure waveform captures the experimentally measured waveform.
	
\section{Conclusion}
 	This investigation sought to quantify the uncertainties in estimations of nonlinear arterial compliance parameters without imposing a particular distribution for the errors of our proposed method.  Use of kernel density estimation in a bootstrapping Monte Carlo procedure allowed such analysis and provided evidence on the stability of our method, both in converging to a narrow set of parameters and in producing waveforms that closely match measured data. As the pressure waveforms used in this study did not contain significant distortions of the diastolic region, it is worth investigating the performance of the proposed parameter estimation method under additional physiological conditions that may contain such distortions.

\bibliographystyle{IEEEtran}
\bibliography{article}
\end{document}